\documentstyle[12pt]{article}

\begin{document}
\input epsf \renewcommand{\topfraction}{0.8}
\pagestyle{empty}
\begin{flushright}
{}
\end{flushright}
\vspace*{5mm}
\begin{center}
\Large{\bf Living on a brane}$^*$

\vspace*{1cm} \large{\bf J.A.R. Cembranos, }{\bf A. Dobado}
{\bf and A. L. Maroto} \\
\vspace{0.2cm}
\normalsize Departamento de  F\'{\i}sica Te\'orica,\\
 Universidad Complutense de
  Madrid, 28040 Madrid, Spain\\
\vspace{0.2cm} \vspace*{0.6cm} {\bf ABSTRACT} \\ \end{center}
\vspace*{5mm} \noindent

We briefly review  the distinctive
signals of brane world models with low tension. We pay special
attention to the brane fluctuations (branons), whose
phenomenological consequences could be important both in high energy
particle physics experiments and in astrophysical and cosmological 
observations.


\begin{flushleft}

\end{flushleft}
\vspace{4cm}

\noindent 
\rule[.1in]{8cm}{.002in}

\noindent $^*$Proceedings of the 29th Bienal Real Sociedad
Espa\~nola de F\'{\i}sica on the
centennial of the society, Madrid, Spain, July 2003.
\vfill\eject

\setcounter{page}{1} \pagestyle{plain}


\newpage

\section{Introduction}

All the  present experimental evidence is compatible with the
existence of only three spatial  and one time dimensions. However,
some  of the  modern theories of the fundamental interactions
predict the existence of additional spatial dimensions. Such
theories include Kaluza-Klein models,  supergravity, string theory
and the more recent M-theory.

In particular, in the last years, it has been found that certain
fundamental extended object, similar to multidimensional membranes
embedded in a higher dimensional bulk space, and known as branes,
play a fundamental role in the dynamics of string and M theory.
Recently, it has been suggested that all the Standard Model (SM)
fields could be confined in one of those (world) branes with three
spatial dimensions whereas gravitation would have access to the
whole bulk space \cite{ADD}. In this scenario, the fundamental
scale of gravity is not the Planck scale, but another scale $M_F$
which could be as low as the electroweak scale. In such a model,
known as ADD model,  the enormous difference between the Planck
scale (10$^{16}$ TeV) and the electroweak one (1 TeV) could be
explained due to the existence of two or more compactified large
extra dimensions (with sizes ranging from 0.1 mm to 1 fm for two
and seven extra dimensions respectively).

\section{New phenomenlogy: KK gravitons, brane fluctuations
and topological states}

Probably the main phenomenological consequences of this scenario
are, on one hand, the modification of  the gravitational
interaction at short distances \cite{ADD}. Thus, whereas at large
distances the inverse square law is respected, it is expected to
be modified at distances close to  the size of the extra
dimensions. On the other hand, the existence of extra dimensions
implies that new fields will appear on the brane, corresponding to
the tower of light Kaluza Klein (KK) modes of the bulk gravitons.

In addition to these features, the fact that a relativistic brane
cannot be a rigid object, requires the existence of some other new
particles describing  its fluctuations with respect to the
equilibrium position \cite{Sundrum}. These excitations, known as
branons, couple to the energy-momentum tensor of the SM fields
with a strength suppressed by the brane tension scale ($f$)
\cite{DoMa}. Another interesting consequence of the brane
flexibility is the suppression \cite{GB} of the couplings of the
KK modes to the SM particles when the tension scale is much
smaller than $M_F$. In this case, for flexible enough branes, the
only relevant degrees of freedom at low energies would be the
branons and the SM fields. Finally, in addition to the branons,
the brane can support also a new set of topological states. These
states are defects that appear due to the non-trivial homotopies
of the vacuum manifold. In fact, string, monopole, skyrmion and
wrapped brane configurations have been studied in different works
\cite{Shifman,BSky}. However, the existence of this kind of states
depends very much on the topology of the extra space, unlike the
brane fluctuations, whose low-energy dynamics is universal,
depending only on the number of branons and the brane tension. The
geometry of the bulk space determines only  the branon masses
\cite{BSky}.

\section{The branon field}

Since the mechanism responsible for the creation of the brane is
in principle unknown, we will assume that the brane dynamics can
be described by an effective action. In this sense, the branons
could be interpreted as the mass eigenstates of the brane
fluctuations in the extra-space directions. Thus, branons are a
kind of new scalar fields, which are massless only in ideal cases
in which the extra space isometries are not explicitly broken, but
only spontaneously broken by the presence of the brane. In such a
case they could be identified with the Goldstone bosons of this
ideal symmetry breaking pattern \cite{Sundrum,DoMa}.

\section{Brane dark matter}

Branons interact typically by pairs. Therefore they are expected
to be stable, massive and difficult to detect  since their
interactions are suppressed by the  tension scale (f). Thus the
massive oscillations of the brane are natural candidates to dark
matter \cite{CDM}  in the brane-world scenario where $f < M_F$
(see FIG. 1).

\begin{figure}[h]
{\epsfxsize=12.0 cm \epsfbox{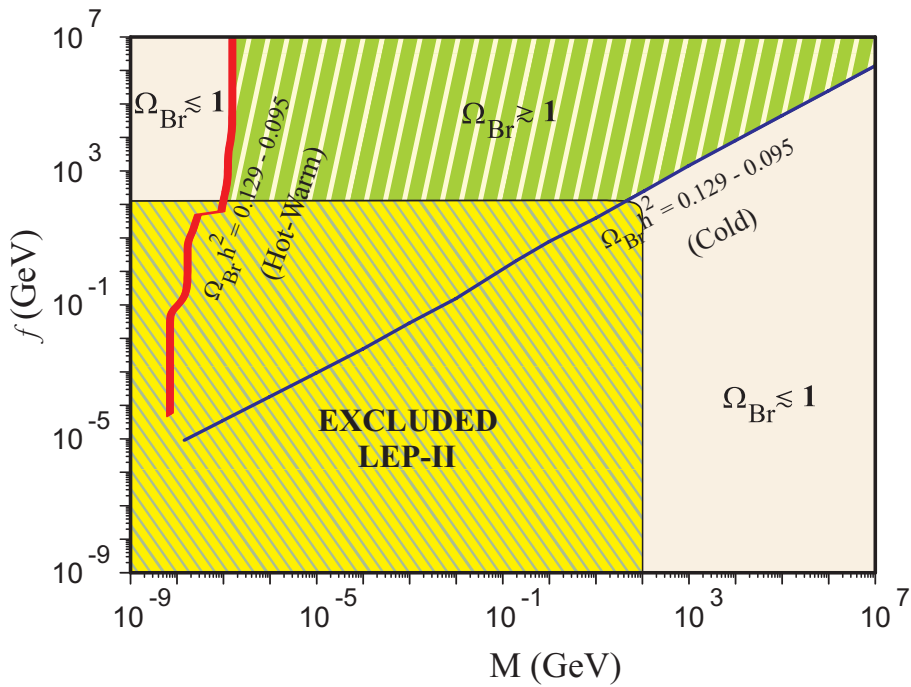}
 \caption{\label{fig1}
Relic abundance in the $f-M$ plane for a model with one branon of
mass: $M$. The line on the left is the $\Omega_{Br}h^2=0.129 -
0.095$ curve for hot-warm relics, whereas the right line
corresponds to cold relics. The lower striped area is excluded by
single-photon processes at LEP-II \cite{ACDM} and the upper area
is also excluded by branon overproduction.}
}
\end{figure}

\section{Future prospects}

Brane fluctuations could be, not only  natural cosmological dark
matter candidates, but they could also make up the galactic halo
and explain the local dynamics. In such case, they could be
detected in direct search experiments through their interactions
with nucleons, or in indirect ones, due to annihilations in the
galactic halo which give rise to pairs of photons,
positron-electron or neutrinos. This kind of experimental analysis
is complementary to the searches in colliders \cite{CrSt,ACDM}
whose present bounds  are also plotted in FIG. 1.

\vspace{.5cm}

{\bf Acknowledgements:} This work
 has been partially supported by the DGICYT (Spain) under
 project numbers FPA2000-0956 and BFM2000-1326.

\end{document}